\documentstyle[aps]{revtex}
\input{epsf.sty}
\begin{document}
\title{Momentum Distribution Spectroscopy Using Deep Inelastic Neutron 
Scattering}
\author{G.F.Reiter}
\address{Physics Department, University of Houston}
\author{J. Mayers}
\address{ Rutherford -Appleton Laboratory, Didcot, England}
\author{J. Noreland}
\address{Defence Research Establishment, Sweden}
\maketitle
\begin{abstract}
We show that deep inelastic neutron scattering from hydrogen(or other light 
nuclei) can be used to measure a spectrum of anharmonic contributions to the 
target atom  momentum distribution with high and known accuracy . The 
method is 
applied here to determine the momentum distribution of the hydrogen in the 
hydrogen bonded system $KHC_2O_4$(potassium binoxalate), where  13 anharmonic 
coefficients 
are 
obtained at 
the 2$\sigma$ to 3$\sigma$ level. The momentum distribution  is  
obtained to an accuracy of better than  few percent at all significant values 
of momentum. 
\end{abstract}
\input{psfig.sty}
\section{Introduction}
The measurement of proton momentum distributions by neutron scattering is
 analogous to the measurement of electron momentum distributions by Compton 
 scattering\cite{pss}  and measurement of  nucleon momenta by Deep Inelastic 
 Scattering \cite{is} and is known as Neutron Compton Scattering  (NCS) or Deep 
 Inelastic Neutron Scattering (DINS).
All three techniques rely upon the fact that if the momentum transferred 
 from the incident to target particle is sufficiently large, the impulse 
approximation (IA) can be 
 used to interpret the data. In the IA, momentum and kinetic energy are 
 conserved. From a measurement of the momentum and energy change of the 
 neutron, the momentum of the target nucleus before the collision can be 
determined.

DINS  measurements  have only become practical since the construction 
 of intense accelerator based neutron sources, which have allowed 
 inelastic neutron scattering measurements with energy transfers in the eV 
 region\cite{hmt}.     Energy transfers 
much 
 greater than  the maximum vibrational frequency of the target atom are required 
before the IA 
 can be used to reliably determine the momentum distribution . At lower 
 energy transfers the IA is no longer valid and  is not related in a simple 
 way to the observed scattering intensities.
 
    There have been a few 
 pioneering studies on anisotropic systems at eV energy 
transfers\cite{rw,ph,isns,pal,ftm}
 but the analysis has been limited to fitting Gaussians to the observed 
 data,  or more generally fitting the data with  model containing a few 
parameters, as was done for measurements on molecular hydrogen\cite{jm1}. We 
show here that an entire spectrum of anharmonic coefficients can be measured 
without recourse to any model, in addition to the 
widths of an anisotropic gaussian, thus  describing an arbitrary anisotropic and 
anharmonic momentum distribution,  
The possibility of doing this for isotropic systems was first suggested by 
Reiter and Silver,\cite{rs} 
 That possibility, for more general systems,  is now a reality. We demonstrate 
this
by measuring the momentum 
distribution for $KHC_2O_4$  where we obtain  14  anharmonic coefficients whose 
size varies by nearly two orders of magnitude, with at least  2-3$\sigma$ 
confidence 
levels for all but one.   The experimental instrument is  the EVS spectrometer 
at ISIS. The work 
presented here by no means represents the limits of resolution of the 
instrument, but rather the first experiments of this kind. Upgrades are planned 
in the near future that will significantly increase flux and counting 
efficiency.
\section*{Theory of Measurement}
The theoretical basis of neutron Compton scattering is the impulse 
 approximation  (IA), which is exact when the momentum transfer and energy 
 transfer   are infinite\cite{n,s,jm2}. The neutron scattering function 
S($\vec{q},\omega$), is related to the momentum distribution n($\vec{p}$) 
in the impulse approximation limit by the relation 

\begin{equation}
S(\vec{q},\omega) = {M\over q}\int n(\vec{p})\delta(y-\vec{p}.\hat q)d\vec{p} = 
{M\over q} J(\hat {q}, y)
\label{sqw}
\end{equation}
where y=${M\over q}(\omega-{q^2\over 2M})$, M is the mass of the target 
particle,q=$|\vec{q}|$, and $\hat{q}$=$\vec{q}/q$.  

DINS measurements on protons have a particularly simple interpretation, as 
 the interaction of protons with other atoms can usually be accurately 
 accounted for\cite{l,hrp,wls} in terms of a single particle potential and hence 
by a 
 proton wave function.\cite{temp} From elementary quantum mechanics, 
n($\vec{p}$)  is 
related to 
 the Fourier transform of the proton wave function   via,
\begin{equation}
n(\vec{p})={1\over(2\pi)^3)}|\int|\Psi(r)exp(i\vec p\cdot\vec r)d\vec r|^2
\label{npsi}
 \end{equation}				
and a DINS  measurement of  n($\vec{p}$) can  be used to determine the wave 
function in 
 an analogous way to the determination of real space structure from a 
 diffraction pattern. If  n($\vec{p}$) is known, and if the proton is in a site 
with reflection symmetry, so that the wave function can be assumed real, then in 
principle both the proton wave 
 function and the exact form of the potential energy well in which the 
 proton sits can be directly reconstructed.\cite{rs}  With an asymmetric site 
such as potassium binoxalate, the phase information that is lost by taking the 
absolute 
value of the momentum wavefunction  is irrecoverable, and we will not be able to 
reconstruct the potential directly. The  n($\vec{p}$) obtained can, of course, 
be used to check any model potential.  

While the original formulation of the inversion problem\cite{rs} is complete as 
it stands, it is useful 
for the systems we will be dealing with to take into account the anisotropy of 
the system explicitly.   
 The fundamental result
that allows for a simple inversion of the Radon transform, J($\hat q, y$) to 
obtain n($\vec{p}$) makes use of a basis of Hermite polynomials and spherical 
harmonics in which the 
transform is diagonal. That is, a single term in the series for J($\hat q, y$) 
corresponds to a single term in the expansion of n($\vec{p}$).

 If we express J($\hat q, y$) in this basis as

\begin{equation}
J(\hat q, y) = {e^{-y^2}\over \pi^{1\over 2}} \sum\limits_{n,l,m} a_{n,l,m} 
H_{2n+l}(y)Y_{lm}(\hat q)
\label{exp}
 \end{equation}

 then n($\vec{p}$) is given in the related basis of Laguerre polynomials as
 
 \begin{equation}
 \label{inv}
 n(\vec{p}) = {e^{-p^2}\over \pi^{3\over 2}}\sum\limits_{n,l,m}2^{2n+l}n!(-1)^n 
a_{n,l,m} p^lL_n^{l+{1\over 2}}(p^2) Y_{lm}(\hat p)
 \end{equation}
 
 where $\hat p$ and $\hat q$ are unit vectors. Clearly, since the expansions are 
complete, a distribution of the form
\begin{equation}
 \label{anh}
 n(\vec p) = \prod\limits_i {e^{-p_i^2\over 2\sigma_i^2}\over 
(2\pi\sigma_i)^{1\over 2}} R(\vec p)
\end{equation} 
with the $\sigma_i$  significantly different from each other, could be expanded 
in this form, 
but even if R($\vec p$) were 1, it would require a large number of terms in the 
series. To avoid this, we show that the anisotropy may be taken into account by 
a change of variables, so that the coefficients $a_{n,l,m}$ represent genuinely 
anharmonic contributions. 

Introducing the new variables

\begin{equation}
p_i^\prime  = p_i /\sqrt2\sigma_i
\end{equation}

with n($\vec p$) defined as in Eq. (\ref{anh}) , defining $R^\prime(\vec 
{p^\prime}$)=R($\vec p(\vec {p^\prime}$)) and

\begin{equation} 
\label{trans1}
n^\prime(\vec p^\prime)={e^{-{p^\prime}^2}\over \pi^{3\over 2}}R^\prime(\vec 
p^\prime)
\end{equation}

 we have

\begin{equation}
\label{trans}
J(\vec{q}, y) =  \int n^\prime(\vec {p^\prime}) \delta(y-\vec {p^\prime}.\vec 
{q^\prime})d\vec {p^\prime} 
\end{equation}
                                                                                                                                                                                                                                                                                                                                     
where ${{q^\prime}_i} = q_i\sqrt2\sigma_i$. The right hand side of Eq. 
(\ref{trans}) is no longer a Radon transform, since $\vec q^\prime$ is not a 
unit vector. However, defining ${y^\prime} = y/\vert\vec q^\prime|$ we obtain
\begin{equation}
\label{fin2}
J(\vec{q}, y) = {1\over \vert\vec q^\prime\vert}\int n^\prime(\vec {p^\prime}) 
\delta(y^\prime-\vec {p^\prime}.\hat {q^\prime})d\vec {p^\prime} = {1\over 
\vert\vec {q^\prime}\vert}J^\prime(\hat {q^\prime}, y^\prime)
\end{equation}
where $J^\prime(\hat {q^\prime}, y^\prime)$ is the Radon transform of the 
isotropic(in it's gaussian component) but anharmonic distribution 
$n^\prime(\vec {p^\prime})$. If $ {\hat q}$ is specified as a unit vector in the 
usual 
spherical coordinates, then 
\begin{equation}
\label{fin}
\vert\vec q^\prime\vert = \sqrt2\big((\sigma_1 sin(\theta)cos(\phi))^2+(\sigma_2 
sin(\theta)sin(\phi))^2+(\sigma_3 cos(\theta))^2\big)^{1\over 2}
\end{equation}
Our procedure is to expand $J^\prime(\hat {q^\prime}, y^\prime)$ in hermite 
polynomials, as in Eq. (\ref{exp}), and least squares  fit the data, 
S($\vec{q},\omega$), using Eqs. (\ref{sqw},\ref{fin2},\ref{fin}), to obtain the 
parameters, $\sigma_i, a_{n,l,m}$.  $n^\prime(\vec {p^\prime})$ can then be 
reconstructed as in Eq. (\ref{inv}), and we thus obtain n($\vec{p}$) as in 
Eq.(\ref{anh}) with R($\vec p$)= $R^\prime(\vec {p^\prime}(\vec p))$. That this 
is a practical procedure will be demonstrated below.
\section{Measurements}
The measurements were performed on the electron volt spectrometer eVS\cite{me}
 at the ISIS neutron source.   On EVS the energy of the scattered neutron is 
 fixed by a resonance filter difference technique\cite{stb}. The final neutron 
 velocity and energy are related by $E_1=m{\nu_1}^2/2$  where m  is the neutron 
mass.  The energy 
 of the incident neutron is determined from a measurement of the neutron 
 time of flight  via the equation
\begin{equation}
\label{tof}
t={L_0\over\nu_0}+{L_1\over\nu_1}
\end{equation}									
where t  is the measured time of flight $,L_0$   and $L_1$  are the lengths of 
the 
 incident and the scattered flight paths of the neutron, $\nu_0$  and $\nu_1$  
are the 
 speeds of the incident and scattered neutrons.  Then
\begin{equation}
\label{endef}
\omega=m({\nu_0}^2-{\nu_1}^2)/2
\end{equation}										
and
\begin{equation}
\label{qdef}
q=m({\nu_0}^2+{\nu_1}^2+2\nu_1\nu_0cos\theta)^{1\over 2}
\end{equation}								
where $\theta$  is the scattering angle.  From  these equations $\omega$   and 
$\vec q$ can be 
 determined for a given time of flight t , if the instrumental parameters 
$L_0,L_1,\theta$   
   and $E_1$ are known. Hence from the count rate at a given time t,  J($\hat q, 
y$) can be determined.  
 On eVS the detectors are situated in the horizontal plane and hence $\vec q$ is 
 always horizontal. By orienting the sample with a chosen crystal axis 
 vertical, it is possible to measure J($\hat q, y$)   for $\vec q$  in whichever 
plane, relative to the sample,  one chooses.
 A time of flight scan at a particular angle for a given detector does not 
correspond, however,
 to a particular direction of $\vec q$. There is significant curvature 
 of this scan through the proton momentum space since the direction of  $\vec q$ 
 varies significantly over the data region. Time of flight spectra for eight 
adjacent detectors at angles between 35 and 55 degrees scan through the atomic 
momentum momentum space of the proton as illustrated in figure 1
\begin{figure}[h]
\hspace{.2 \hsize}
\psfig{figure=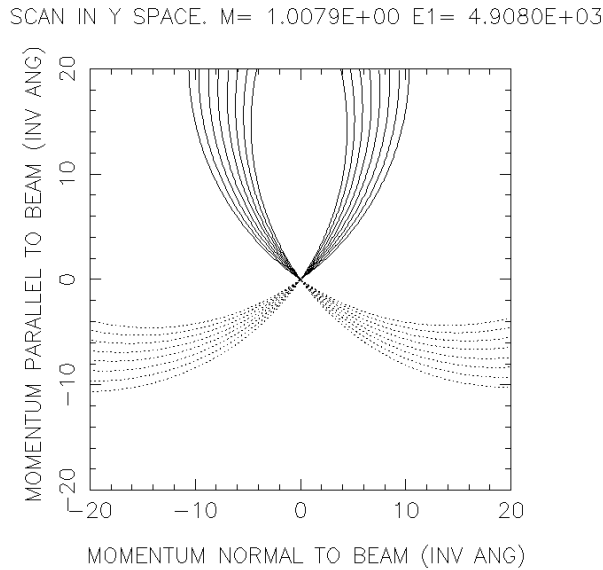,width=100mm}

\caption{Scan pattern in momentum space for detectors at a fixed angle as the 
time of flight is varied }
\label{fig1}
\end{figure}

A complete scan over the proton momentum space is constructed by combining a 
 number of data sets, taken with the sample rotated about the vertical axis 
 by appropriately chosen angles.

The reported measurements were made using  two banks of 8 $ Li^6$ doped glass 
 scintillator detectors which were symmetrically placed on each side of the 
incident beam at scattering angles 
 between  $35^o$ and $55^o$. For DINS studies of protons it is necessary to 
 site the detectors at forward scattering angles since the hydrogen 
 scattering cross section is strongly anisotropic at eV incident energies, 
 with virtually no back scattering.  This restriction is a kinematic 
 consequence of the closeness of the mass of the neutron and the hydrogen 
 atom and does not apply to heavier atoms.

The resolution function  of the instrument is determined by the 
 uncertainties in the measured values of the time of flight t and the 
 distribution of $L_0,L_1,\theta$ and  $E_1$ values allowed by the instrumental 
geometry and 
 analyser foil resolution. Uncertainties in $L_0$  arise primarily from the 
finite 
 depth of the neutron moderator,  those in $L_1$  and $\theta$   from the finite 
 sample and 
 detector sizes and those in t  from jitter in the detector electronics. All 
 resolution components can be determined by calibration measurements and all 
 except the energy component can be approximated by Gaussians, without 
significant error. A 0.015 
 mm thick gold foil provided a Lorentzian energy resolution function at 
$E_1=4908meV$ , 
 with a peak HWHM of 136 meV.  The Gaussian and Lorentzian  resolution 
 components in momentum space y ,are listed in table 1 for two angles 
 representative of the range of angles employed. The resolution  is 
 dominated by the energy component which varies strongly with scattering 
 angle. The second most important contribution comes from the angular 
 resolution of the spectrometer and is independent of angle. The momentum 
 and energy transfers at the centre of the hydrogen response peak are also 
 listed for the different angles.
\vskip .25in
{\bf Table 1.} The resolution widths are the  Lorentzian HWHM for  (RL)  and the 
Gaussian standard deviation for other parameters (RG) . The momentum q and 
 energy transfer ( at the scattering angles $35^o$  and $55^o$ are also given.)
\begin{center}
\begin{tabular}{||c|c|c|c|c||}
\hline 
\hline 
 Angle & $R_G(\AA^{-1})$ & $R_L(\AA^{-1})$ & q $(\AA^{-1})$ & $ \omega$ (eV)\\
\hline
\hline 
$ 35^0$    & 0.61       & 1.08 & 34.1 & 2.41\\
\hline
$ 55^0$ & 0.55 & 0.55 & 48.8 & 4.92\\
\hline
\hline
\end{tabular}
\end{center}
The raw data contains signals from all the atoms in the scattering sample 
 and from the cryostat background. Fortunately the energy transfer to 
 hydrogen is much greater than that to other atomic masses and the proton 
 signal is well separated from that due to other masses.   The contribution 
 from all components other than hydrogen is subtracted by fitting a sum of 
 Gaussians convoluted with the instrument resolution function to the data 
 and subtracting off the fitted contribution to other peaks. There is also a 
 small contribution to the data from a second gold resonance at  60 eV which 
 can be seen at ~100 $\mu$sec and this is also fitted and subtracted from the 
 data. 
\begin{figure}[h]
\hspace{.2 \hsize}
\psfig{figure=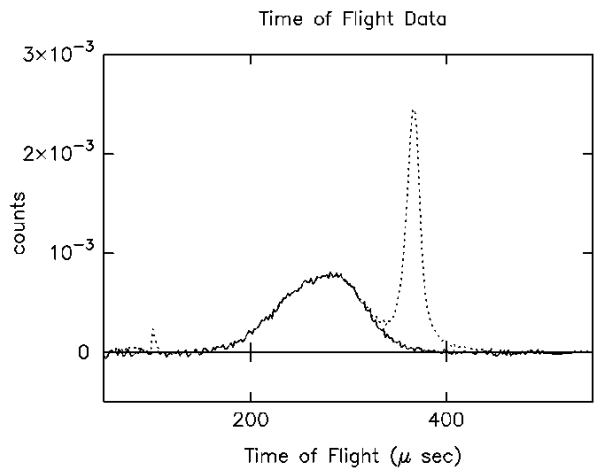,width=100mm}

\caption{ The sum of data from 8 detectors at scattering angles between $38^o$ 
 and $55^o$ is shown  as the dotted line. The data after 
 subtraction of the contribution from atoms with higher masses and the 60 eV 
 resonance data  is shown as the full line.The total data set for a single plane 
 consisted 
of 36 such spectra. The FWHM of the Lorentzian resolution function is ~6 
microsec.}
\label{fig2}
\end{figure}
The data for each scan was converted into a distribution in the momentum 
 space of the crystal as described above. Complete coverage of the plane was 
 achieved by combining six runs which were taken at steps of $23^o$ sample 
 rotation. A contour plot of the data derived from the six runs is shown in 
 figure 3. This was produced by binning the counts from different detectors and 
sample orientations in the appropriate pixel of the momentum space
\begin{figure}[h]
\hspace{.2 \hsize}
\psfig{figure=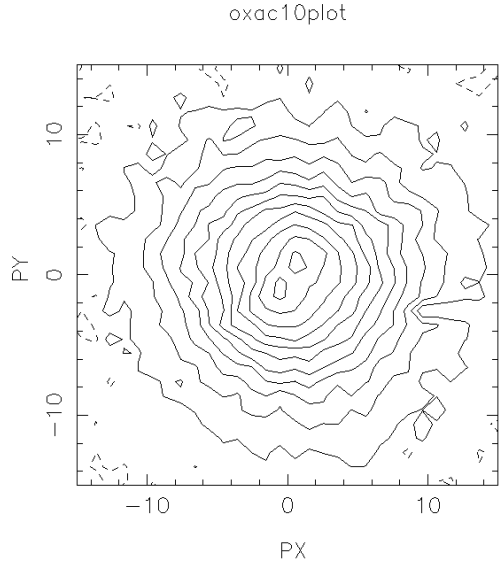,width=100mm}

\caption{Contour plot of data in one plane of potassium binoxalate. The apparent 
double 
peak near the origin is an artifact of the method of plotting the data  }
\label{fig3}
\end{figure}

The data has been corrected for sample attenuation, but still contains 
 errors due to small deviations from the Impulse Approximation which are 
 present at the finite momentum transfers of the measurement. These tend to 
 introduce small asymmetries into the data set at the ~1-5\% level, thereby 
 removing the exact inversion symmetry of the Compton profile. It has been 
 shown by Sears\cite{s2} that most of these effects are removed by 
 symmetrisation of the data about the origin. This procedure cannot be 
 followed for our data sets as the scan in a single detector is curved and 
 different points in the crystal plane which are related by inversion 
 symmetry may have been measured in different detectors, with different 
 experimental resolution. However by fitting  to a J($\hat q, y$  with inversion 
symmetry, 
 as discussed below,  we automatically include a correction for  deviations from 
the impulse 
 approximation. Any asymmetries are ignored by the fit and should not 
 affect the values of the fitted parameters.

\section{Fitting Procedures}
Eqs.(\ref{exp},\ref{inv}) hold quite generally for Radon transform pairs, but 
physical requirements in the present context restrict the allowed coefficients. 
Since 
$J(\vec{q}, y)$ is an even function of y, l is restricted to even values, and 
since  $J(\vec{q}, y)$is real, the $a_{n,l,m}$ for $\pm m$ must be equal. In 
fact, if there are residual final state effects in the data, the data will not 
be symmetric. Restricting the coefficients in this way will therefore eliminate 
these residual effects. Lifting the restrictions and fitting the data allows one 
to measure the extent of these effects.For potassium binoxalate, data was taken 
for  
three perpendicular planes oriented parallel to the crystal axes The procedure 
followed was to 
 perform a simultaneous fit to the 6x16x3=288 separate time of flight spectra 
 to an expansion of the form given in Eq.(\ref{exp}), convoluted with the 
instrument resolution function
  
The actual fitting procedure requires that three of the parameters $\sigma_i, 
a_{n,l,m}$ be fixed, as they are not all independent. That is, if one varies the 
$\sigma_i$ abitrarily, there will always be a set of $a_{n,l,m}$ that will fit 
the data, the particular values that fit depending on the choice of the 
$\sigma_i$. One could obtain fixed values for the $\sigma_i$ by first fitting 
the data to a gaussian, i.e. $R^\prime(\vec p^\prime)=1$, and then varying only 
the $a_{n,l,m}$, in which case all such terms would in principle be needed. We 
find in practice, that better results are obtained, i.e. fewer anharmonic 
coefficients needed and more rapid convergence of the series for $R^\prime(\vec 
p^\prime)$, when the $\sigma_i$ are allowed to vary but the first three 
anharmonic coefficients, $a_{1,0,0}, a_{0,2,0}, a_{0,2,2}$ are set to zero. In 
fact, these coefficients are not really anharmonic coefficients at all. They 
could always be eliminated by a shift of the  $\sigma_i$ and an adjustment of 
higher order coefficients. This procedure thus has the virtue as well of 
producing only genuinely  anharmonic corrections to a gaussian fit. 

The datasets as they are presently obtained in the EVS spectrometer at ISIS, are 
obtained one plane at a time(see discussion of experimental apparatus). That is 
$\vec q$ varies within a plane, and a range of y values is taken such that 
$J(\vec{q}, y) $is negligible outside this range. 
There is a very high density of points, which for the purposes of the present 
discussion we can take to be continuous.  The question then arises, 'How many 
planes of data
are needed to determine a specified number of coefficients, and at what angles 
to each other should they be?' We can see the point of the question by 
considering first a single plane and looking at the fit to the leading 
anharmonic coefficient. From Eq.(\ref{exp}) we see that there are six 
independent coefficients  multiplying $H_4$, i.e. the coefficients of 
$(Y_{00},Y_{20}, Y_{22}, Y_{40}, Y_{42}, Y_{44})$.  Let us say that our 
coordinate system is chosen so that the plane measured is treated as an xz 
plane. Then in terms of the variable $\theta$ there are only three independent 
fourier coefficients that can be present in the data for the coefficients of 
$H_4$, i.e. the coefficients of (1, $cos(2\theta), cos(4\theta)$ for instance. 
Therefore three of the coefficients are not independent. The complete set of 
coefficients cannot be determined by the data. This is of course due to the fact 
that there is no information as to the behaviour of $J(\vec{q}, y)$ for nonzero 
azimuthal angles $\phi$  in the data, so we shouldn't expect the fitting 
procedure to provide it.  The data provides a complete description of  
$J(\vec{q}, y)$ only if this function is rotationally invariant about the z 
axis. Actually, what is required is only that $J^\prime(\hat {q^\prime}, 
y^\prime)$ be rotationally invariant, since the fit is done in the primed 
coordinate system. If this is the case, then all coefficients of $Y_{lm}$ with 
non-zero values of m must be zero. We see that there are only three remaining 
possibly non zero coefficents, which can all be determined. For higher order 
terms as well, keeping only the coefficients with m=0 provides all the 
independent terms needed to fit the data, and the resulting fit, of course, is 
rotationally symmetric about the z axis.

If the data is not known to be rotationally symmetric, additional planes of data 
must be taken to determine even these lowest order coefficients. In general, 
whenever we take another plane of data, we might expect to obtain 3 more 
independent measurements of the coefficients of $H_4$, 4 independent 
measurements of the coefficients of $H_6$,
and in general, k+1 measurements of the coefficients of  $H_{2k}$. (k+1 being 
the number of independent fourier components in the data for that value of k). 
Since the number of  $a_{n,l,m}$ that are to be determined for 2n+l=2k is 
(k+2)(k+1)/2, it appears that (k+2)/2 planes are needed to measure all 
coefficients up to $H_{2k}$. The angles between the planes must be chosen, 
however, so that the measurements are really independent. For instance, if k=2, 
it would appear that two planes would suffice, but if they are chosen as the xz 
and yz planes, they do not provide independent measurements of the coefficients. 
 This may be seen by observing that the sum of the data from the two planes 
gives three independent fourier coefficients to determine four independent  
$a_{n,l,m}$, the coefficients of $(Y_{00},Y_{20}, Y_{40}, Y_{44})$, since the 
coefficients of  $Y_{22}$ and  $Y_{42}$ cannot affect this sum. The difference 
of the data on the two planes gives three equations for the two coefficients of 
$Y_{22}$ and  $Y_{42}$.  A better choice for the planes would be $\phi=0$ and 
$\phi=\pi/4$, which would allow the determination of all the coefficients. If 
there is some symmetry in the problem, one may be able to use perpendicular 
planes  if the symmetry axis is chosen appropriately with respect to the common 
axis of the two scattering planes. For instance, if there is tetragonal 
symmetry present, and the symmetry axis is chosen perpendicular to the common 
axis, one can obtain all the allowed coefficients up to k=4. One can show that 
three perpendicular planes do in fact suffice to determine the coefficients up 
to k=4 in the general case, without any symmetry to reduce the number of 
allowed coefficients. The question of whether this is an optimum configuration 
of planes or not, we will leave to another time. Including the three  
$\sigma_i$, a three plane measurement allows 34 coefficients to be measured. 
 
\section{Measurement Errors}
The uncertainty in the measurement of n($\vec p$) at some point $\vec p$ is due 
to the uncertainty in the measured coefficients. Denoting an arbitrary 
coefficient by $\rho_i$, we have 
\begin{equation}
\label{uncn}
\delta n(\vec{p})= \sum\limits_{i} {\delta n(\vec{p})\over \delta\rho_i} 
\delta\rho_i
\end{equation}
 The fitting program, after a minimum is obtained with some set of coefficients, 
 calculates the correlation matrix $<\delta\rho_i\delta\rho_j>$ by varying the 
coefficients slightly and calculating the curvature of the chi-square of the 
fit. \cite{dev}. Hence, the variance in the momentum distribution is

\begin{equation}
\label{deln}
<\delta n(\vec{p})^2>= \sum\limits_{i,j} {\delta n(\vec{p})\over \delta\rho_i} 
{\delta n(\vec{p})\over \delta\rho_j}<\delta\rho_i\delta\rho_j>
\end{equation}
There are of course, potential systematic errors that could enter the 
measurement, such as multiple scattering effects,  or an error in determining
the resolution function. The former must be handled with good experimental 
design, and if small, can be corrected for. We have done measurements on samples 
whose thickness differed by a factor of two, with no significant differences in 
the observed scattering. Multiple scattering effects would also lead to 
asymmetries in J($\hat q, y$) that are not observed. The resolution function 
has been studied extensively, and is believed to be known
 accurately.In the present measurement, the resolution width is between 15 and 
25\% of the width of the distribution measured.

 A further source of error not 
contained in the estimate in Eq.(\ref{deln}) is the truncation of the series 
used to fit the data to include 
only terms up to 2n+l=8. To get an idea of the seriousness of this, and to test 
the fitting procedures and software, we have generated synthetic data from a 
known momentum distribution that corresponds to an asymmetric double well, and 
convolved it with the instrumental resolution function. The data was then 
analyzed by the means described above, and the extracted n($\vec{p}$)  compared 
with the input. The input  n($\vec{p}$) corresponded to a spatial wave function 
consisting of two displaced gaussians with the same variance and a relative 
weight r=.5. The explicit form is 
\begin{equation}
\label{dpotn}
n(p_x,p_y,p_z)={(1+r^2+2rcos(2p_za))\over{ (1+r^2+2re^{-{a^2\over 
2{\sigma_z}^2}}})}
 \prod\limits_i {e^{-{p_i^2\over 2\sigma_i^2}}\over (2\pi\sigma_i)^{1\over 2}}
\end{equation}

where $\sigma_x=4.6, \sigma_y=4.0,\sigma_z=6.0 $ and a=.15 in units of inverse 
angstroms and angstroms, respectively. This form is rather similar to the actual 
form of the data we will analyze. The coordinate system z axis was chosen 
to be identical with the crystal z axis.  The comparison is shown in fig 4. The 
extracted  n($\vec{p}$) is plotted with a dotted line, the input n($\vec{p}$) 
above with a dashed line. As may be seen from the figure, there is essentially 
no error due to the truncation of the  the expansion. Of course, an input with 
more variation might require including higher terms in the expansion, and hence 
taking more planes of data. The actual data we have obtained for potassium 
binoxalate has 
less variation than the simulation and the truncation error will be negligible. 
In the other directions, the fit is rigorously gaussian, since all coefficients 
with $m\not=0$  were zero. We note that the measured $\sigma_z$ parameter was 
4.47, not 6.0. It is simply a fitting parameter. The overall momentum 
distribution has physical significance, the individual parameters may not. In 
fact, we have gotten the same degree of fit with the crystal z axis aligned 
along the coordinate y axis. In this case, all the coefficients are non-zero, 
but the resultant n($\vec{p}$) is identical to the one displayed above.  
\begin{figure}[h]
\hspace{.2 \hsize}

\psfig{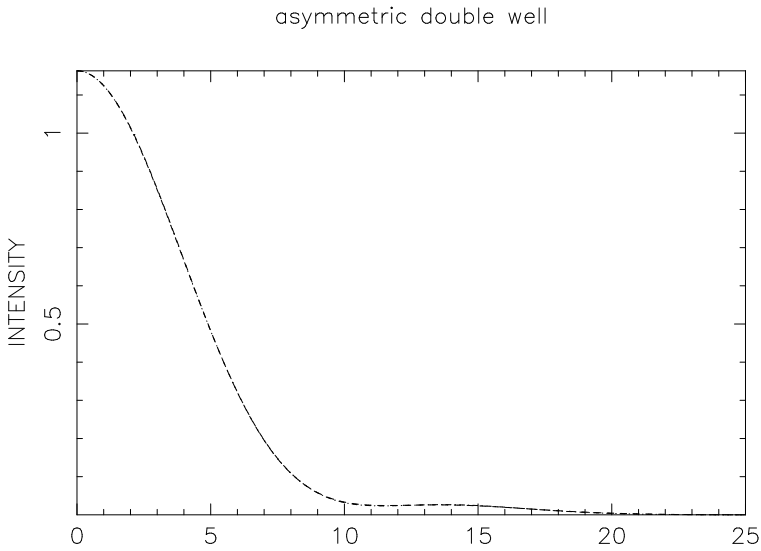}

\caption{Comparison of input n($\vec{p}$), given by Eq.(16)with reconstruction 
of n($\vec{p}$ )using fitting procedure. The z coordinate axis is chosen to be 
the double well axis.} 
\label{fig4}
\end{figure}
A final source of systematic error involves the possibility of finding a false 
minimum with the fitting procedure. With such a large number of parameters, 
there is the possibility that the program will home in on a local minimum and 
miss the true minimum. We are using straightforward gradient methods for the 
search, which would have difficulty with a very rugged chi-squared landscape. We 
do not appear to have such a landscape for these problems, and certainly not for 
the large parameters, such as the $\sigma_i$, or the largest anharmonic 
coefficients, whose values appear to be quite robust with regards to different 
paths to the minimum. One can check for this problem by orienting the fitting 
coordinate system differently with respect to the crystal axes, as was done for 
the test data above.  In this case, all 
the coefficients will be different, but the final n($\vec{p})$ must be the same.  
There is also the fact that  n($\vec{p})$ must be positive, 
and a spurious fit that leads to significant negative values can be rejected. We 
can also eliminate some parameters which are consistently much smaller than 
their variance,and which can be set to zero without affecting significantly the 
chi-square, thus  reducing the dimensionality of the space. We believe there are 
no problems of this sort with the fits we will present.  
\section{Results for Potassium Binoxalate}
Potassium Binoxalate, $KHC_2O_4$,  is a hydrogen bonded system in 
which the hydrogen sits in an asymmetric position between two oxygen 
atoms.\cite{emd}.The crystal is monoclinic. We will choose a primitive cell for 
which the bond axis, that is the line joining the two oxygen atoms, 
is essentially aligned with the c axis of the crystal.\cite{cell} We choose this 
axis for 
the z axis of our coordinate system. Three planes of data were taken at right 
angles to each other, with 69,632 data points in all. One of these is the bc 
plane, where the b axis is the unique axis, and the other two are the $a^*b$ and 
$a^*c$ plane, where the $a^*$ axis is perpendicular to the bc plane.  These were 
fit with the 
methods described above to give the momentum distribution. The measurements were 
done at 10 deg K and hence there are no significant finite temperature 
corrections to the ground state momentum distribution due to excited states. 

 Note that the coordinate system used to describe the momentum distribution need 
not have the symmetry of the crystal. The local symmetry of the site is only a 
two-fold  rotation.  We show the values of 
the fitted coefficients in tables  2 and 3, along with the rms uncertainty in 
their 
values. This is included only to give some sense of the individual parameter. 
The error in n($\vec{p}$)is given by Eq.(\ref{uncn}) and includes 
the 
effect of the correlations between coefficients, whereas the figures cited in 
the 
table are only the diagonal correlation coefficients. It can be seen, that many 
of 
the measured coefficients have been determined at the 2-3 $\sigma$ level of 
confidence, some at higher levels, and only one at a 1$\sigma$ level.   
Coefficients that are set to zero in the fitting procedure were 
found to have values smaller than their variance by at least a factor of 2, and 
setting them to zero did not significantly change the minimum value of 
chi-square.  The 
goodness of fit to a sample of the data is shown in Fig. 5.
\begin{figure}[h]
\hspace{.2 \hsize}
\psfig{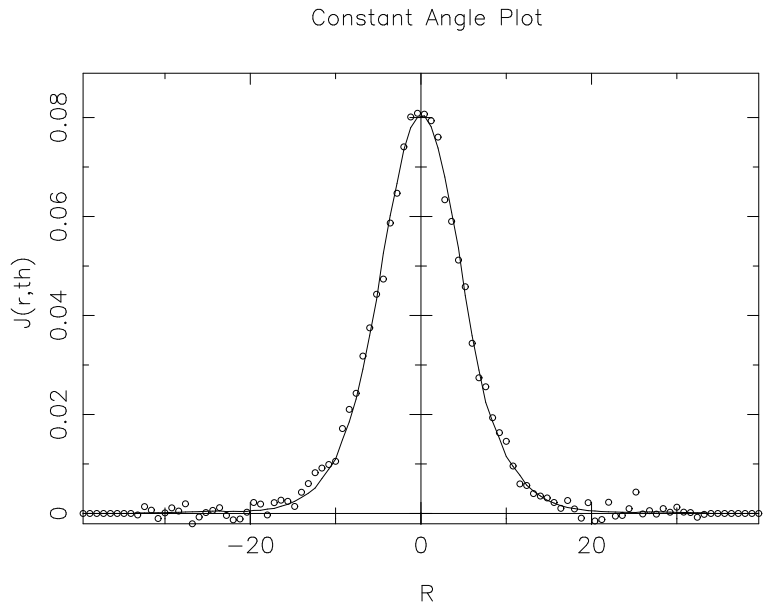}

\caption{Data fitted by method described above. The data is a composite of data 
points in a 10deg wedge about the z axis. 
}
\label{fig5}
\end{figure}

 We have compared the 
fitted prediction, convolved with the instrumental resolution, and the data, for 
the cumulative sum of data in a 10 degree wedge along the hydrogen bond 
direction. The measured 
momentum distribution along the  axes of the measurement planes, are shown in 
Figs. 6-9, and 
contour plots along coordinate planes in Figs. 10-13.

\begin{figure}[h]
\hspace{.2 \hsize}
\psfig{figure=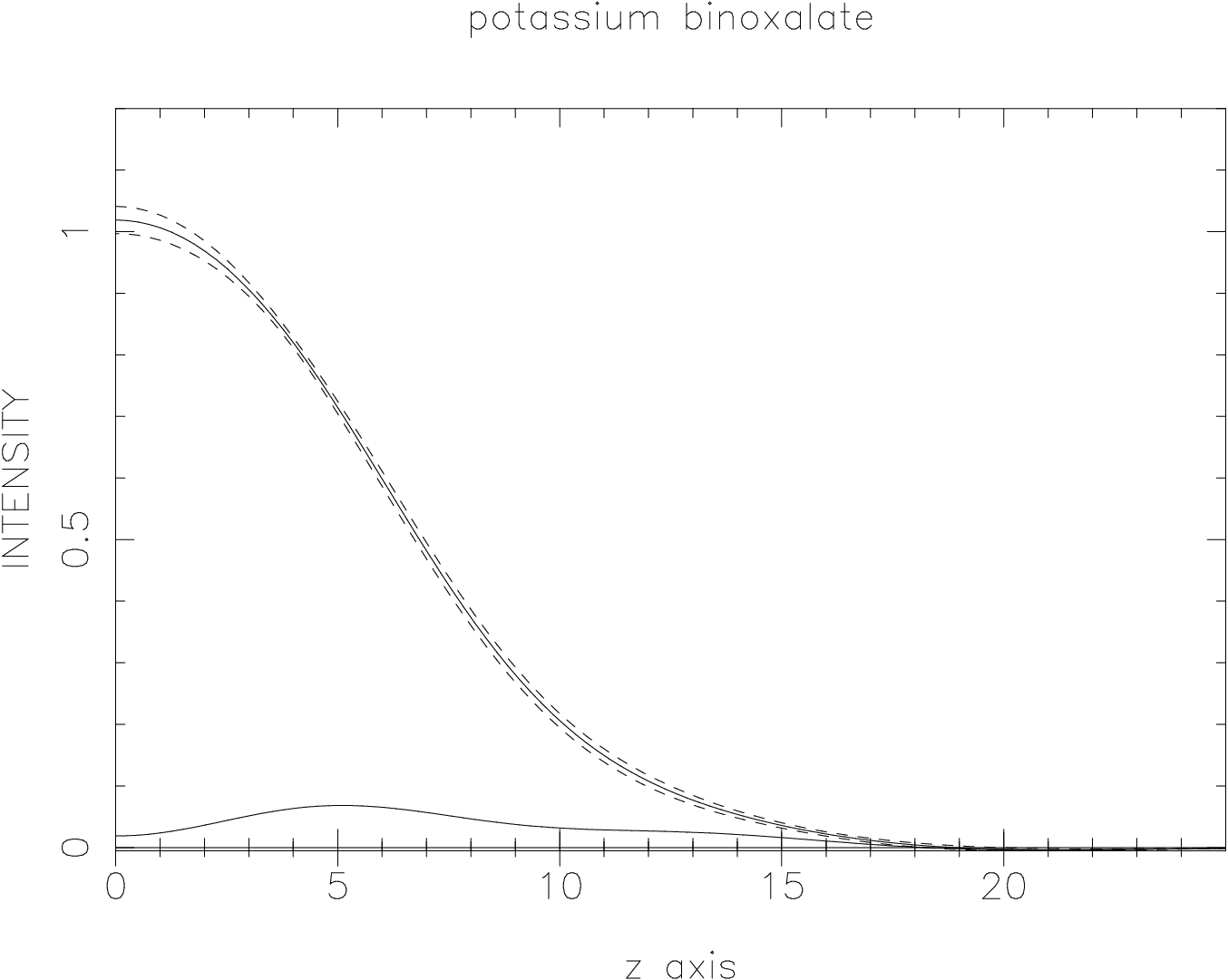,width=100mm,angle=-90}
\caption{Momentum distribution for potassium binoxalate along the axis shown. 
The 
momentum is in units of $\AA^-1$, n($\vec{p}$ is in arbitrary units. The errors 
are calculated as in Eq.(15) for the parameters that are significant in Tables 2 
and 3. The lower curve is the anharmonic component.
}
\label{fig6}
\end{figure}
\begin{figure}[h]
\hspace{.2 \hsize}
\psfig{figure=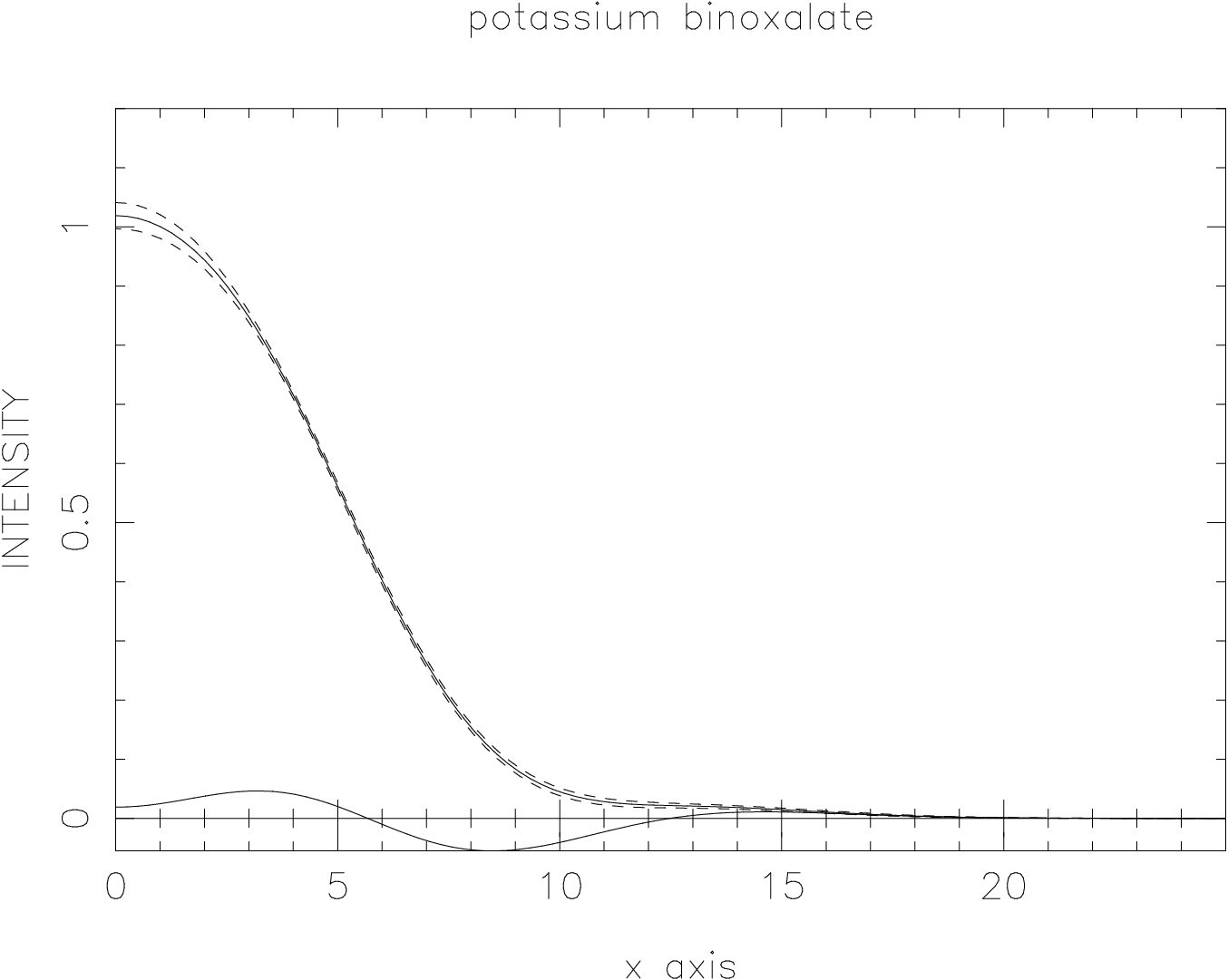,width=100mm,angle=-90}

\caption{Momentum distribution for potassium binoxalate along the axis shown. 
The 
momentum is in units of $\AA^-1$, n($\vec{p}$) is in arbitrary units. The errors 
are calculated as in Eq.(15) for the parameters that are significant in Tables 2 
and 3. The lower curve is the anharmonic component.
}
\label{fig7}
\end{figure}
\begin{figure}[h]
\hspace{.2 \hsize}
\psfig{figure=pgy.ps,width=100mm,angle=-90}

\caption{Momentum distribution for potassium binoxalate along the axis shown. 
The 
momentum is in units of $\AA^{-1}$, n($\vec{p}$) is in arbitrary units. The 
errors are calculated as in Eq.(15) for the parameters that are significant in 
Tables 2 and 3 }. The lower curve is the anharmonic contribution. 
\label{fig8}
\end{figure}
\begin{figure}[h]
\hspace{.2 \hsize}
\psfig{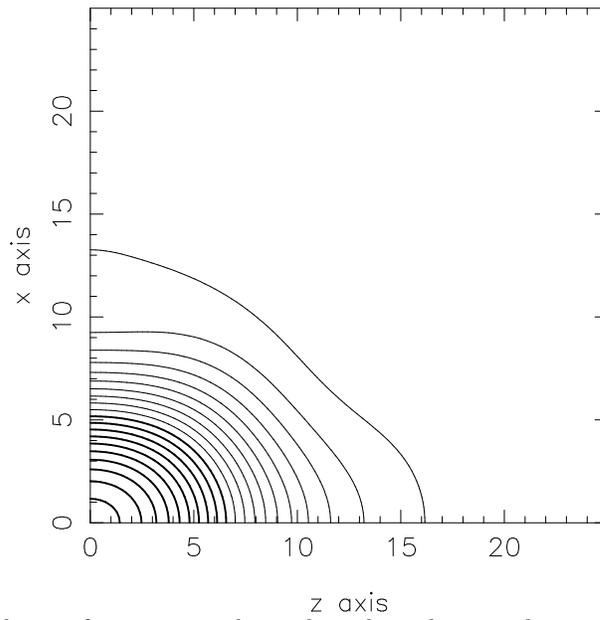}

\caption{Momentum distribution for potassium binoxalate along the axes shown. 
The 
momentum is in units of $\AA^{-1}$, n($\vec{p}$) is in arbitrary units.  }
\label{fig9}
\end{figure}
\begin{figure}[h]
\hspace{.2 \hsize}
\psfig{figure=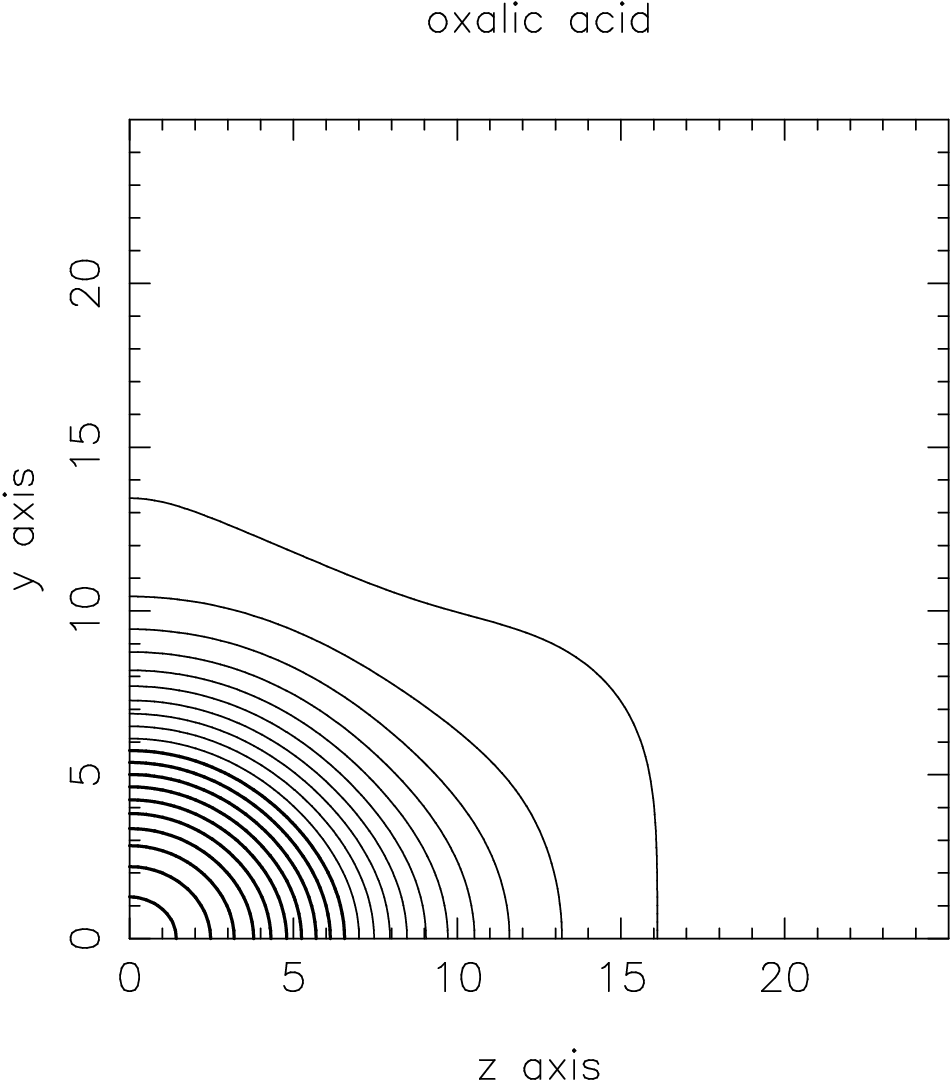,width=80mm,angle=-90}

\caption{Momentum distribution for potassium binoxalate along the axes shown. 
The 
momentum is in units of $\AA^{-1}$ }
\label{fig10}
\end{figure}

\begin{figure}[h]
\hspace{.2 \hsize}
\psfig{figure=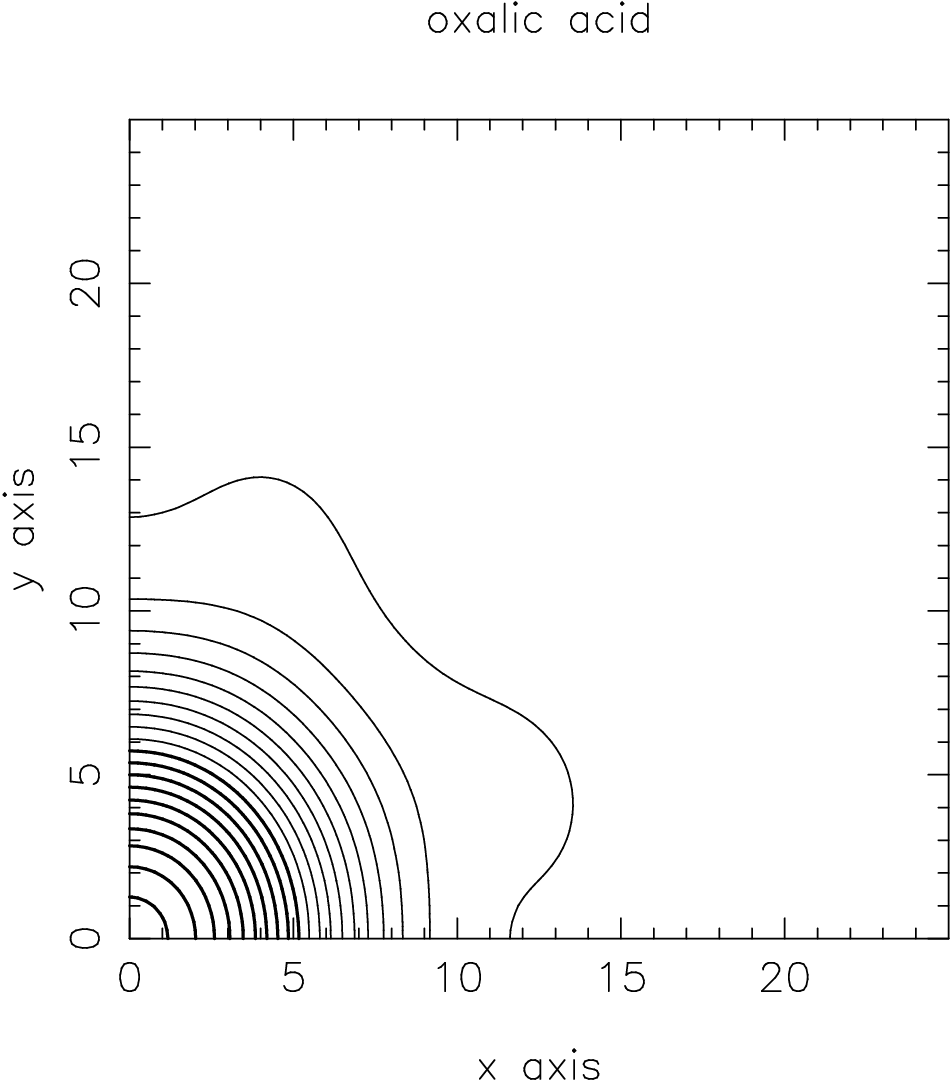,width=80mm,angle=-90}

\caption{Momentum distribution for potassium binoxalate along the axes shown. 
The 
momentum is in units of $\AA^{-1}$. }
\label{fig11}
\end{figure}
\newpage
\begin{figure}[h]
\hspace{.2 \hsize}
\psfig{figure=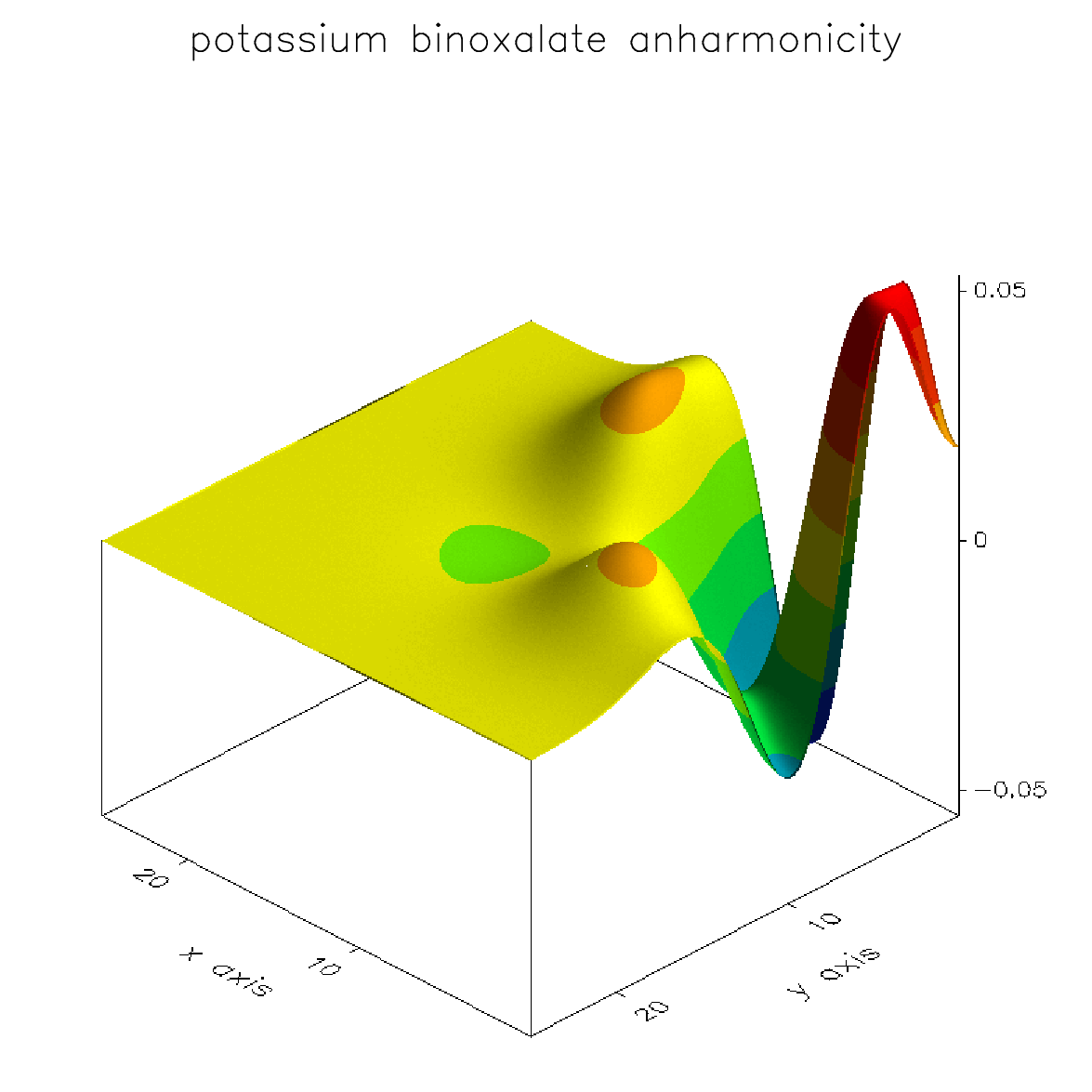,width=80mm,angle=-90}
\caption{Anharmonic contribution to the momentum distribution for potassium 
binoxalate along the axes shown. The 
momentum is in units of $\AA^{-1}$. }
\label{fig12}
\end{figure}

\begin{figure}[h]
\hspace{.2 \hsize}
\psfig{figure=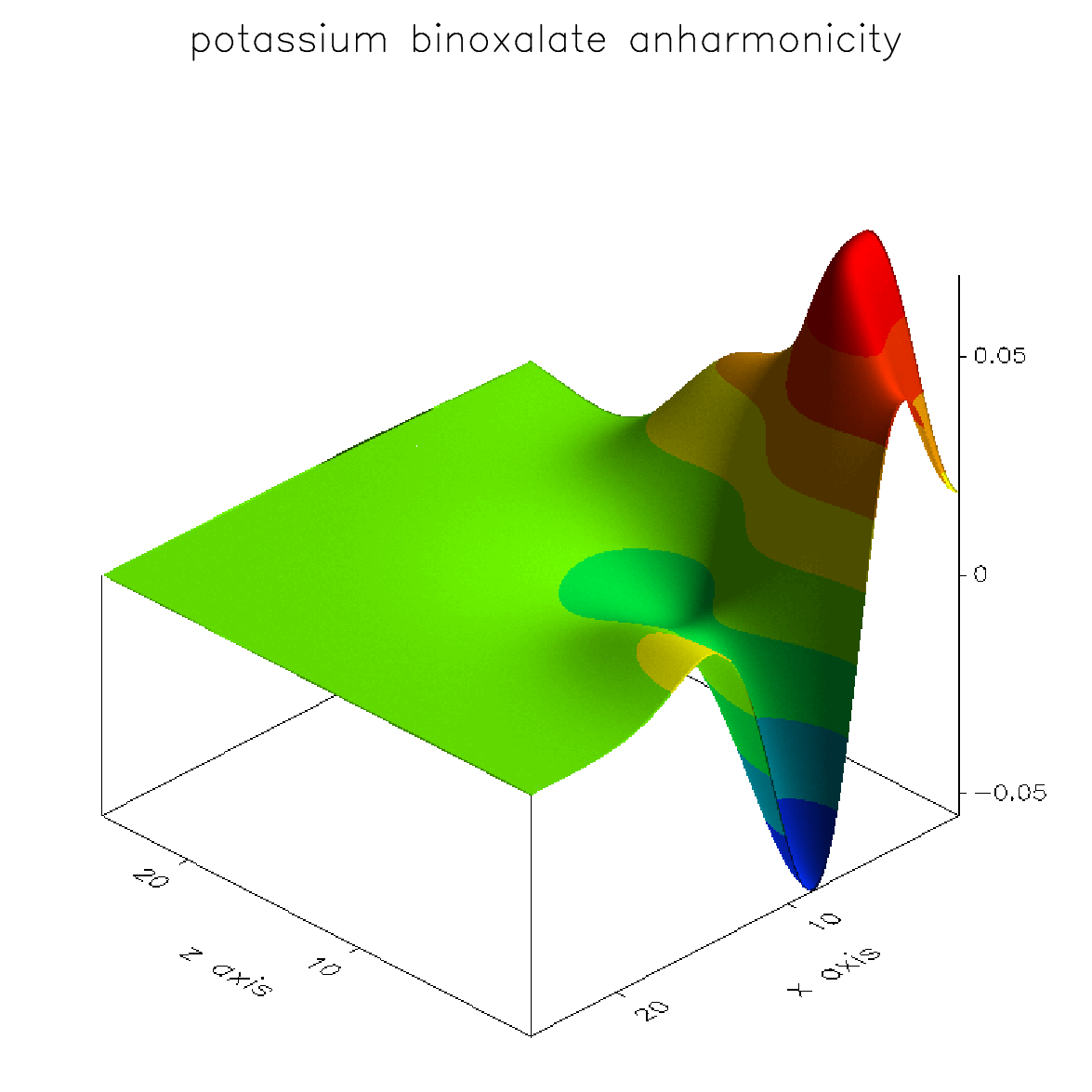,width=80mm,angle=-90}
\caption{Anharmonic contribution to the momentum distribution for potassium 
binoxalate along the axes shown. The 
momentum is in units of $\AA^{-1}$. }
\label{fig13}
\end{figure}
\newpage

The coefficients that were measured are given in the tables below.
\vskip .25in
{\bf Table 2.} The harmonic fitting coefficients and the variances in their 
values as measured
\begin{center}

\begin{tabular}{||c|p{.6in}|p{.6in}||}
\hline
\multicolumn{3}{||c||}{\bf Harmonic Coefficients}\\
\hline
i&$\sigma_i$&$\delta\sigma_i$\\
\hline
x&4.509&.0235\\
y&4.869&.0254\\
z&5.351&.0449\\
\hline\hline
\end{tabular} 
\end{center}
\smallskip

\vspace{.5in}
{\bf Table 3.} The anharmonic fitting coefficients and their variances as 
measured.
\begin{center}
\begin{tabular}{||c|c|c|p{.6in}|p{.6in}||}
\hline 
\multicolumn{5}{||c||}{\bf Anharmonic Coefficients}\\
\hline 
 n & l & m & $a^{\prime}_{n,l,m}$ & $\delta a^{\prime}_{n,l,m}$\\
\hline
 2 & 0 & 0 & 0.0000 &0 .0000\\
1 & 2 & 0 & -0.1443&0.0377\\
1&2&2&0.0000 &0.0000 \\
0&4&0&0.0510 &0.0086 \\
0&4&2& -0.0564&0.0095 \\
0&4&4&0.0000 &0.0000 \\
3&0&0&-0.1356 &0.0098 \\
2&2&0&-0.0296 &0.0105 \\
2&2&2&0.0000 &0.0000 \\
1&4&0&-0.000 &0.0000 \\
1&4&2&0.0000 &0.0000 \\
1&4&4&0.0000 &0.0000 \\
0&6&0&-0.0184 &0.0065 \\
0&6&2&-0.0127 &0.0028 \\
0&6&4&-0.0388 &0.0129 \\
0&6&6&0.0000 &0.0000 \\
4&0&0&-0.0290 &0.0029 \\
3&2&0&0.0000 &0.0000 \\
3&2&2&0.0000 &0.0000 \\
2&4&0&-0.0016 &0.0013 \\
2&4&2&0.0069 &0.0013 \\
2&4&4&0.0000 &0.0000 \\
1&6&0&-0.0049 &0.0021 \\
1&6&2&0.0000 &0.0000 \\
1&6&4&-0.0098 &0.0036 \\
1&6&6&0.0000 &0.0000 \\
0&8&0&0.0000 &0.0000 \\
0&8&2&0.0000 &0.0000 \\
0&8&4&0.0000 &0.0000 \\
0&8&6&0.0000 &0.0000 \\
0&8&8&-0.0075 &0.0028 \\
\hline\hline
\end{tabular}
\end{center}
\smallskip

\vskip .1in
Note that there are 14 anharmonic coefficients that are measureable, 13 of which 
are at the $2\sigma$ level at least, and that  the harmonic coefficients are 
measured 
to better than 1\%. 
\section{Conclusion}
We have shown  how the method of analysis of DINS data suggested in \cite{rs}
can be extended to anisotropic momentum distributions, and have applied this 
method to an analysis of the hydrogen bond in potassium binoxalate.   
The results demonstrate that DINS, as it is 
implemented now at ISIS, is capable of  detailed, model independent,  
measurement of the momentum distribution for hydrogen, and by inference, other 
light atoms.
These measurements required about four days of beam time, and  are the first 
such measurements to be analyzed in this way.  The count rates, resolution and 
detetector efficiency can be, and are scheduled to be, significantly improved. 
The data can be analyzed in less than a day. The DINS technique thus provides a 
practical means of accessing precise information about the anharmonicity of 
local potentials and  can provide a check of any theoretical calculation of 
these potentials at a level of accuracy and detail that has not been possible 
previously.   
\section{Acknowledgements}
We would like to thank Devinder Sivia for assistance in analyzing the data, 
Steve Bennington and John Tomkinson for useful discussions, and Professors 
R.G.Delaplane and H. Kuppers for providing the crystals.

\appendix
\section{Seperable Distributions}
The distribution used for the simulations described here is a special case of a 
general separable distribution
 \begin{equation}
 \label{ap1}
 n(\vec{p}) = \prod\limits_i n_i(p_i)
 \end{equation} 
If we represent 
\begin{equation}
 \label{ap2}
 n_i( p_i) = {e^{-p_i^2\over 2\sigma_i^2}\over{(2\pi\sigma_i)^{1\over 2}}} 
\sum\limits_n a_{i,n}H_n({p_i\over{\sqrt2\sigma_i}})
\end{equation}
make use of the fact that 
\begin{equation}
 \label{ap3}
 \int e^{iqx-x^2}H_n(x)dx=\pi^{1\over 2}e^{-q^2\over 4}(iq)^n,
\end{equation}
and represent the delta function in the definition of the radon transform  by 
its fourier transform, it is straightforward to show that the Radon transform of 
$n(\vec{p})$ in the isotropic coordinate system defined in the text is given by 
\begin{equation}
J^\prime(\hat {q^\prime}, y^\prime)= {e^{-y^\prime 2}\over{\pi^{1\over 2}}}  
\sum\limits_{n_j}( \prod\limits_{i} a_{i,n_j}({{q_i^\prime }\over{ 
q^\prime}})^{n_j}) H_{n_1+n_2+n_3}(y^\prime)
\label{ap4}
\end{equation}
For the case we have used as a simulation, in which the distribution is 
anharmonic only in the z direction, the result simplifies to
\begin{equation}
J^\prime(\hat {q^\prime}, y^\prime)= {e^{-y^\prime 2}\over{\pi^{1\over 2}}}  
\sum\limits_{n}( a_{3,n}(cos(\theta^\prime)^n) H_{n}(y^\prime)
\label{ap5}
\end{equation}
In this case, the coefficients in the expansion of  $J^\prime(\hat {q^\prime}, 
y^\prime)$ in the form given by Eq.(\ref{exp}are not all independent. There is 
only one independent coefficient in the set of 
 $a_{n,l,0}$ for each value of 2n+l.


\begin{references}
\bibitem{pss} P. M. Platzmann in `Momentum Distributions', ed R. N. Silver  and 
P. E.Sokol (Plenum Press,New York, 1989, p249)
\bibitem{is} I Sick in ref 1, page 175.
\bibitem{hmt} R. S. Holt, J. Mayers and A. D. Taylor  in ref 1 , p. 295
\bibitem{rw}  H. Rauh and N. Watanabe, Phys. Lett. 100A, 244 (1984)
\bibitem{ph} M. P. Paoli and R. S. Holt, J. Phys. C 21, 3633 (1988)
\bibitem{isns} S. Ikeda, K. Shibata, Y. Nakai amd P. W. Stephens, J. Phys. Soc. 
Japan 61,2619 (1992)
\bibitem{pal} P. Postorino, F. Fillaux, J. Mayers, J. Tomkinson and R.S. Holt,
J. Chem. Phys. 94, 4411 (1992)
\bibitem{ftm} A L Fielding, D N Timms and J Mayers Europhys. Lett. 44 255 (1998)
\bibitem{jm1} J. Mayers, Phys. Rev. Letts. 71, 1553, (1993)
\bibitem{rs} R. Silver and G.reiter, Phys.Rev.Letts.{\bf 54},1047,(1985)
\bibitem{n} R Newton, 'Scattering Theory of Waves and Particles', (Springer, 
Berlin,1981)
\bibitem{s} V. F. Sears  Phys.  Rev.  B  30, 44 (1984)
\bibitem{jm2} J. Mayers Phys. Rev. B 41,41 (1991)
\bibitem{l} S. W. Lovesey, ` Theory of Neutron Scattering from Condensed 
Matter',(Oxford University Press, New York,1987)
\bibitem{hrp} R. Hempelmann, D. Richter and D. L. Price Phys. Rev. Lett. 58, 
1016
\bibitem{wls} M. Warner, S. W. Lovesey and J. Smith Z. Phys. B 39, 2022 (1989)
\bibitem{temp}This assumes that the temperature is sufficiently low that the 
proton is essentially in its ground state. This condition is well satisfied for 
the experiments we will discuss.
\bibitem{me}  J. Mayers and A. C. Evans, Rutherford Laboratory Report,  
RAL-91-048 (1991)
\bibitem{stb} P A  Seeger, A D Taylor and R M Brugger Nuc Inst Meth A240 98 
(1985)
\bibitem{s2} V. F. Sears Phys Rev 185 200, (1969), Phys Rev A7 340 (1973)
\bibitem{dev} D. Sivia, 'Data Analysis:A Bayesian Tutorial', Clarendon Press, 
Oxford (1996)

\bibitem{emd}H. Einspahr,R.E. Marsh and J. Donohue, Acta. Cryst.{\bf B28}, 
2194,(1972)
\bibitem{cell} This choice differs from the choice in the previous reference, 
and corresponds to choice 3 for the unit cell of crystal type 14 with the b axis 
the unique axis in the standard 
tables, International Tables for Crystallography, Vol. A, P 177, D. 
Riedel,(1983),  
\end{references}
\end{document}